\documentclass{aa}
\usepackage{amssymb}
\usepackage{amsmath}
\usepackage{graphicx}
\usepackage{enumerate}   
\usepackage{color,xcolor}
\usepackage{txfonts}
\usepackage{natbib}
\usepackage[colorlinks=true,citecolor=blue,urlcolor=blue,linkcolor=blue]{hyperref}

\def\gr{$\gamma$-ray}

\begin{document}

\title{Account of the baryonic feedback effect in \gr\ measurements of intergalactic magnetic fields}

\author{K.~Bondarenko\inst{\ref{i1},\ref{i2}} \and A.~Boyarsky\inst{\ref{i3}} \and A.~Korochkin\inst{\ref{i4},\ref{i5},\ref{i6}} \and A.~Neronov\inst{\ref{i4},\ref{i7}} \and D.~Semikoz\inst{\ref{i4},\ref{i5},\ref{i8}}\and A.~Sokolenko\inst{\ref{i9}}}

\institute{Theoretical Physics Department, CERN, Geneva 23, CH-1211, Switzerland\label{i1} \and L’Ecole polytechnique federale de Lausanne, 1015 Lausanne, Switzerland\label{i2} \and Institute Lorentz, Leiden  University, Niels Bohrweg 2, Leiden, NL-2333 CA, the Netherlands\label{i3} \and APC, Universite Paris Diderot, CNRS/IN2P3, CEA/IRFU \email{alexander.korochkin@apc.in2p3.fr}\label{i4} \and Institute for Nuclear Research of the Russian Academy of Sciences, 60th October Anniversary st. 7a, 117312, Moscow, Russia\label{i5} \and Novosibirsk State University, Pirogova 2, Novosibirsk, 630090 Russia\label{i6} \and Astronomy Department, University of Geneva, Ch. d'Ecogia 16, 1290, Versoix, Switzerland\label{i7} \and National Research Nuclear University MEPhI, 115409 Moscow, Russia\label{i8} \and Institute of High Energy Physics, Austrian Academy of Sciences, Nikolsdorfergasse 18, 1050 Vienna, Austria\label{i9}}

\authorrunning{Bondarenko et al.}
\titlerunning{Account of the baryonic feedback effect in \gr\ measurements of IGMFs}
\abstract
{}
{
Intergalactic magnetic fields in the voids of the large-scale structure can be probed via measurements of secondary \gr\ emission from \gr\ interactions with extragalactic background light. Lower bounds on the magnetic field in the voids were derived from the nondetection of this emission. It is not clear a priori what kind of magnetic field is responsible for the suppression of the secondary \gr\ flux: a cosmological magnetic field that might be filling the voids, or the field spread by galactic winds driven by star formation and active galactic nuclei.
}
{  
We used IllustrisTNG cosmological simulations to study the effect of magnetized galactic wind bubbles on the secondary \gr\ flux. 
}
{
We show that within the IllustrisTNG model of baryonic feedback, galactic wind bubbles typically provide energy-independent secondary flux suppression at a level of about 10\%. The observed flux suppression effect has to be due to the cosmological magnetic field in the voids. This might not be the case for the special case when the primary \gr\ source has a hard intrinsic \gr\ spectrum that peaks in the energy range above 50~TeV. In this case, the observational data may be strongly affected by the magnetized bubble that is blown by the source host galaxy.
}
{}


\maketitle

\section{Introduction}

Magnetic fields in galaxies and clusters are thought to result from an amplification of initial seed fields that might be of primordial origin (see, e.g.,~\citet{Durrer:2013pga} for a review). This possibility suggests that measuring the parameters of the seed fields can provide a new cosmological observable, a new ``window'' on the early Universe, more specifically, on the epochs predating the epoch of formation of the cosmic microwave background signal. Large-scale magnetic fields that fill the volume of the voids of the large-scale structure (LSS) likely did not experience adiabatic contraction or dynamo amplification, which transformed the seed fields in galaxies and galaxy clusters. The properties of these volume-filling intergalactic magnetic fields (IGMF) might therefore be close to the properties of the cosmological seed fields. 

The technique of measuring the IGMF that is best suited for probing magnetic fields in the voids is based on \gr\ observations of distant active galactic nuclei (AGN). The technique exploits the effect of attenuation of very high-energy (VHE) \gr\ flux from distant AGN by the effect of pair production on photons of the extragalactic background light (EBL). This leads to the development of an electromagnetic cascade along the line of sight. The cascade emission is observable in the form of magnetic field-dependent delayed~\citet{plaga95} or extended \citet{Neronov:2007zz,Neronov:2009gh} emission around the primary AGN point source. Searches for extended and delayed \gr\ emission in the energy band of the Fermi Large Area Telescope (LAT) have not yielded measurements of the void IGMF, but have imposed a lower bound on its strength \citet{Neronov:1900zz,Taylor:2011bn,dermer11}. The most recent analysis of Fermi/LAT data reported by~\citet{Biteau:2018tmv} constrained the IGMF in the voids to be stronger than $\sim 10^{-16}$~G for large correlation-length magnetic fields and $\sim 10^{-14}$~G for short correlation-length fields originating from the early Universe. The next-generation \gr\ observatory, the Cherenkov Telescope Array (CTA), will be able to probe stronger IGMFs with strengths up to $\sim 10^{-12}$~G \citet{Korochkin:2020pvg,Abdalla:2020gea}. 

Even if the CTA will succeed in measuring the IGMF strength in the voids of the LSS, it is not clear a priory whether this field can be identified as the primordial field. The problem is that the voids can be polluted by the magnetic field that is spread by the baryonic feedback process that returns matter from galaxies into the intergalactic medium. The galactic winds driven by supernova and AGN activity are most probably ionized and carry magnetic fields with them \citet{Bertone:2006mr,Pinsonneault:2010zt}. The uncertainties of the details of the baryonic feedback on the LSS do not allow us to answer the question whether the magnetic field that is spread by the feedback can fill the voids and thus hide the primordial field. If this is the case, measuring the IGMF with the CTA will fail to provide a new cosmological observable. Instead, it would rather provide constraints on the baryonic feedback process. 

Magnetized galactic winds can affect the \gr\ measurements of IGMFs only if they produce IGMFs with a volume filling factor on the order of unity \citet{Dolag:2010ni}. The mean free path of \gr s with energies in the 10-30 TeV range is in the 0.1-1 Gpc range. The extended and delayed emission from electrons and positrons deposited in the intergalactic medium by the pair production process is accumulated throughout this length scale. The extended and delayed \gr\ signal is sensitive to the field that spans the largest part of the line of sight toward distant AGN. If the baryonic feedback field is concentrated in the nodes and filaments of the LSS, the delayed and extended \gr\ signal is not strongly affected by the feedback field, and \gr\ measurements can still provide information on the primordial field. 

Significant progress toward a better modeling of the baryonic feedback process has recently been achieved by IllustrisTNG cosmological simulations~\citet{nelson18,springel18,pillepich18,naiman18,marinacci18}. Recently, \citet{Garcia:2020kxm} have used IllustrisTNG to show that AGN and supernova-driven outflows create extended magnetized bubbles with magnetic fields of about $B>10^{-12}$ G and typical sizes in the range of $\sim 10-30$ Mpc, with magnetic fields whose parameters are largely independent of those of the  preexisting seed magnetic fields. 

In what follows, we use the IllustrisTNG model to assess the effect of contamination of the intergalactic medium by the magnetic field that is spread by the galactic winds and its consequences for the measurements of primordial IGMF. As discussed in~\citet{pakmor2017}, the magnetic fields produced by galactic activity are almost insensitive to the initial seed fields. The reason for this is that inside galaxies, the magnetic fields are strongly amplified by dynamo processes, saturate, and ``forget'' their initial conditions. We consider several characteristics situations of VHE \gr\ sources at different distances, with different \gr\ spectra, situated inside and outside the galactic wind bubbles. We find that within the IllustrisTNG baryonic feedback model, the basic features of the extended or delayed \gr\ signal sensitive to the primordial IGMF are just slightly altered, typically by 10-20\%. This does not preclude the possibility of measuring the primordial field in the voids, but it introduces an additional systematic uncertainty that needs to be properly taken into account in the analysis with the future instruments.

\section{Magnetized bubbles in IllustrisTNG simulations}

\subsection{TNG simulations}

IllustrisTNG is a set of gravo-magnetohydrodynamic simulations, the next generation (TNG) of the Illustris project~\citet{nelson18,springel18,pillepich18,naiman18,marinacci18}. It is based on the moving-mesh \textsc{Arepo} code \citet{springel2010MNRAS.401..791S} that solves the system of equations for self-gravity and ideal magnetohydrodynamics~\citet{2011MNRAS.418.1392P,2013MNRAS.432..176P}. 
Cosmological parameters are chosen to be equal to the best-fit Planck 2015 cosmological parameters~\citet{Plank2016A&A...594A..13P}.

We used the highest-resolution simulations that are publicly available TNG100-1 (TNG100)~\citet{Nelson2019ComAC...6....2N}. The TNG100 simulation has a box size $\sim (110~\text{cMpc})^3$ and contains $1820^3$ dark matter particles and an equal number of initial gas cells with masses of $m_{\text{DM}} = 7.5 \times 10^6~M_{\odot}$ and $m_{\text{bar}} = 1.4 \times 10^6~M_{\odot}$. 

The initial seed magnetic field in this run was a constant magnetic field with a field strength $10^{-14}$~cG (comoving Gauss) directed along the $z$ -axis in the simulation box.\footnote{Different configurations of initial magnetic fields were studied in many previous works, see, e.g., \citet{Vazza:2017qge} and reference therein.}This seed magnetic field experienced adiabatic contraction during the structure formation process, and then it was strongly amplified by small-scale dynamos in collapsed structures. 

The TNG simulations adopt a comprehensive galaxy and supermassive black holes (SMBH) formation and feedback model~\citet{Weinberger2017MNRAS.465.3291W,Pillepich2018MNRAS.473.4077P}. SMBHs with an initial mass of $\sim 10^{6}$\,M$_{\odot}$ are placed in the gravitational potential minima of dark matter halos when the virial mass exceeds $\sim 7 \times 10^{10}$\,M$_{\odot}$. These black holes grow via binary mergers with each other or via smooth gas accretion according to the Bondi-Hoyle-Lyttleton model~\citet{2017MNRAS.465.3291W}. The rate of black hole growth depends on the black hole mass, local gas density, and relative velocity between the black hole and its surroundings.

Within the IllustrisTNG model, the SMBH activity has two different regimes that correspond to high-accretion and low-accretion states~\citet{Weinberger2017MNRAS.465.3291W}. In the high-accretion rate state (accretion rate above $\sim 10\%$ of the Eddington limit), the energy is continuously injected into the surrounding gas, causing thermal heating of the gas. In the low-accretion rate state, the kinetic energy is deposited periodically when enough energy accumulates via accretion (see~\citealt{Weinberger2017MNRAS.465.3291W} for additional details). Each injection event creates a randomly oriented high-velocity kinetic wind. If averaged over long time intervals, energy injection in this mode also becomes isotropic. The low accretion rate feedback mode drives the most powerful outflows in the TNG model ~\citet{2019MNRAS.490.3234N}. A large fraction of the injected kinetic energy in this mode thermalizes via shocks in the surrounding gas, thereby providing a distributed heating channel.

\begin{figure}
    \includegraphics[width=\linewidth]{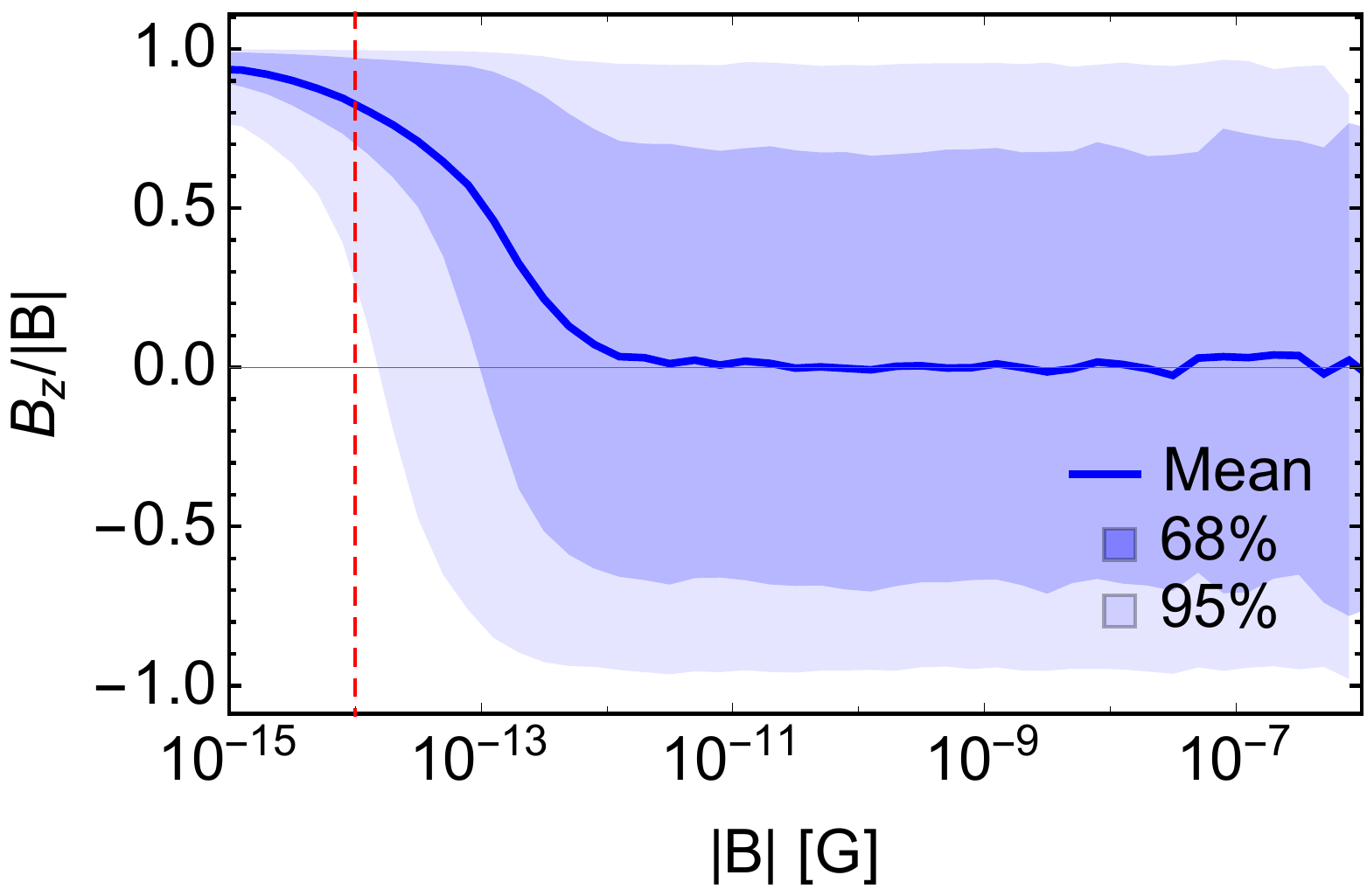}
        \caption{Orientation of the magnetic field along the $z$ direction as a function of the magnetic field strength at redshift $0$. The blue line shows the mean value, while shaded dark and light blue regions represent areas with $68\%$ and $95\%$ of simulation volume. The dashed red line shows the seed magnetic field value $B=10^{-14}$~G.}
        \label{fig:Borientation}
\end{figure}

\subsection{Magnetic bubbles}

According to the TNG simulations, magnetized bubbles (regions larger than the galaxy size with magnetic fields stronger than $10^{-12}$~G, which is orders of magnitude stronger than the initial seed field) of chemically enriched material emerge at $z \lesssim 2$ as a result of  active baryonic feedback from galaxies (see Section 3 in \citealt{Garcia:2020kxm,Garcia:2021cgu} and Fig.~1 therein for the visualization of magnetic bubbles). These bubbles are significantly nonspherical and can have sizes of several dozen megaparsec. \citet{Garcia:2020kxm} demonstrated that at $z=0,$ these bubbles occupy $12-14\%$ of the cosmological volume for magnetic fields stronger than $10^{-12}$~G and more than $3\%$ for $B>10^{-9}$~G. Comparing the simulations with initial magnetic fields that are different by many orders of magnitude (otherwise identical), \citet{Garcia:2020kxm,Garcia:2021cgu}  have shown that magnetic fields in the bubbles are independent of the strength of the initial seed field to a large extent. Fig.~\ref{fig:Borientation} illustrates this effect:  orientations of strong magnetic fields with $B>10^{-12}$~cG ``forgets'' the direction of the seed field (which was along the z-axes in the TNG simulations). In contrast, weak fields in the ``real'' IGM outside the bubbles prefer the initial direction. It was also demonstrated in \citet{Garcia:2020kxm} that the bubbles are mainly caused by the AGN activity (with a smaller contribution by supernovae). We took the magnetic field value $10^{-12}$~cG as a boundary value for the magnetic bubbles.\footnote{In Fig.~1 in~\citet{Garcia:2020kxm},the boundaries of bubbles are quite sharp and there is no significant difference in the volume filling fraction for magnetic fields stronger than $10^{-12}$~cG or $10^{-11}$~cG (see Fig.~4 therein). Therefore the results of this paper do not significantly depend on this specific value.}

\begin{figure*}
    \includegraphics[width=\textwidth,trim={0 1cm 2.5cm 2.5cm},clip]{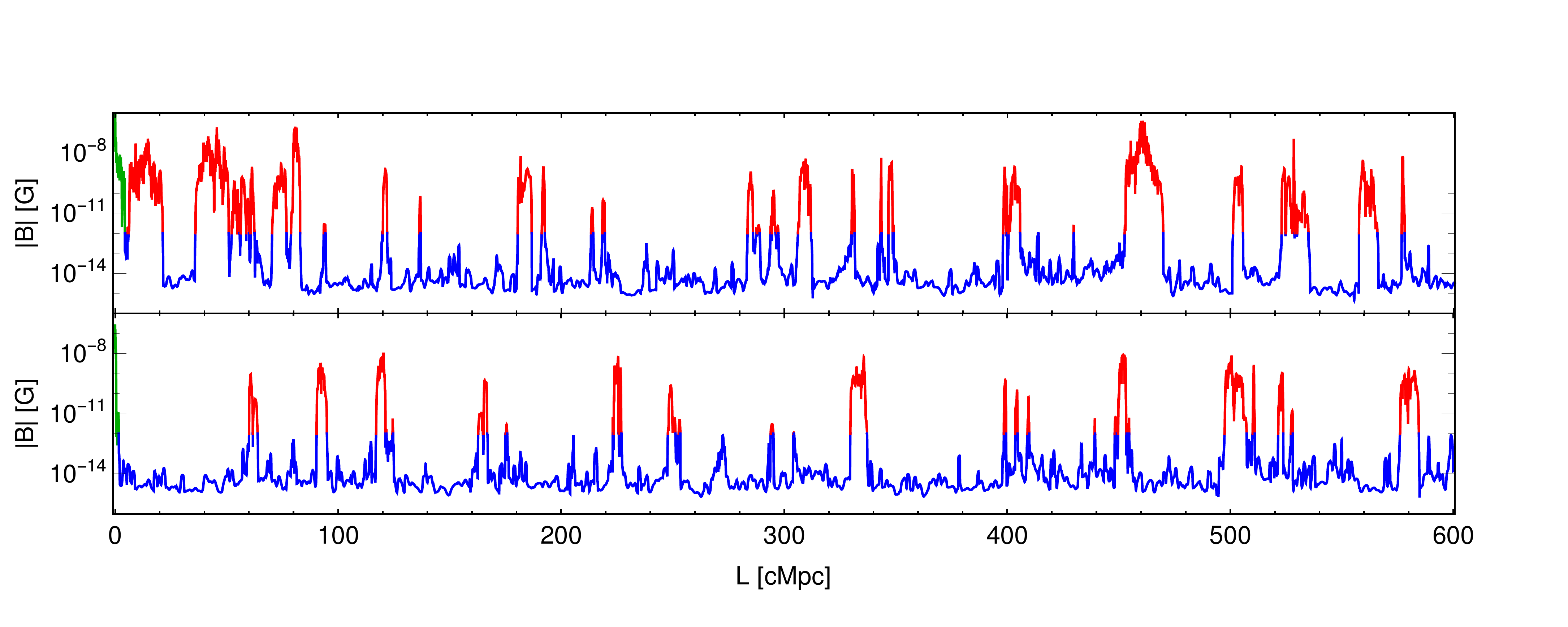}
    \caption{Examples of magnetic field profiles along two 600~Mpc long LOS that start at AGN found in the TNG-100 simulation. The upper panel shows the LOS starting inside a strong magnetic field bubble at the beginning, corresponding to AGN-3. The lower panel shows an example of the LOS starting from AGN-2 within a small bubble. We indicate the first bubble with green, other bubbles with red, and voids with blue. See section~\ref{sec:numerical-modeling} for details.}
    \label{fig:LOSexamples}
\end{figure*}

\subsection{Line-of-sight data}

We are interested in the effects of the bubbles on the extended and delayed \gr\ signal from distant AGN. To assess these effects, we extracted the information on the strength and orientation of the magnetic field along randomly chosen straight lines from the TNG-100 simulation, so-called lines of sight (LOS), originating from seven different SMBHs with masses from $10^{8} M_{\odot}$ to $10^{10} M_{\odot}$. The SMBHs were selected requiring sufficient released energy in the low-accretion rate feedback mode, $E_{\rm low} > 10^{58.5}~\rm{ erg}$. Using continuous boundary conditions in the simulation box, we produced four randomly oriented LOSs with lengths up to 600~Mpc per AGN. Magnetic fields were calculated as average values inside voxels of size $(20\text{ckpc})^3$ using the publicly available pysph-viewer code~\citet{pysph}. Examples of two LOSs starting from AGNs within and outside bubbles with strong magnetic field are shown in Fig.~\ref{fig:LOSexamples}.

\section{Numerical modeling}
\label{sec:numerical-modeling}

To explore the effects of magnetic bubbles on the secondary emission, we numerically modeled the propagation of the high-energy gamma-rays through the cosmic medium using data on a magnetic field for a set of LOSs, extracted from the TNG100 simulation. For this purpose, we used the recently developed Monte Carlo simulation code CRbeam \citet{Berezinsky:2016feh} that takes gamma-ray absorption with pair production on the EBL, subsequent inverse Compton scattering of electrons and positrons on the CMB and EBL, and deflection of the charged component in magnetic fields into account. This code was also tested through a comparison with several alternative cascade codes \citet{Taylor:2011bn,Kalashev:2014xna,Kachelriess:2011bi}. 

In all simulations, the position of the source corresponds to the first voxel of the LOS. The directions of the initial momenta of all particles are the same and coincide with the direction along the LOS. The strength and direction of magnetic fields are given by linear interpolation between neighboring voxels. The secondary \gr\ signal generated by every electron and positron only depends on the strength and orientation of the magnetic field right at the location of electron-positron pair production (more precisely, within the voxel in which the pair production happens). This appears to be reasonable in the view of the fact that electrons with energies $E_\mathrm{e} < 5$~TeV (which produce secondary gamma-rays with $E_\mathrm{\gamma} < 100$~GeV) are typically deflected by $\delta > 3^\circ$ already within the first 20~kpc of their trajectory. Because we are only interested in secondary \gr s that are emitted within the angle that is comparable to the point spread function (PSF) of \gr\ telescopes, $\theta_\mathrm{psf}\lesssim 1^\circ$, further \gr\ production by electrons or positrons that are deflected by more than several degrees does not contribute to the secondary \gr\ signal that is sampled by the telescope situated on the LOS.

\begin{figure*}
    \includegraphics[width=\textwidth]{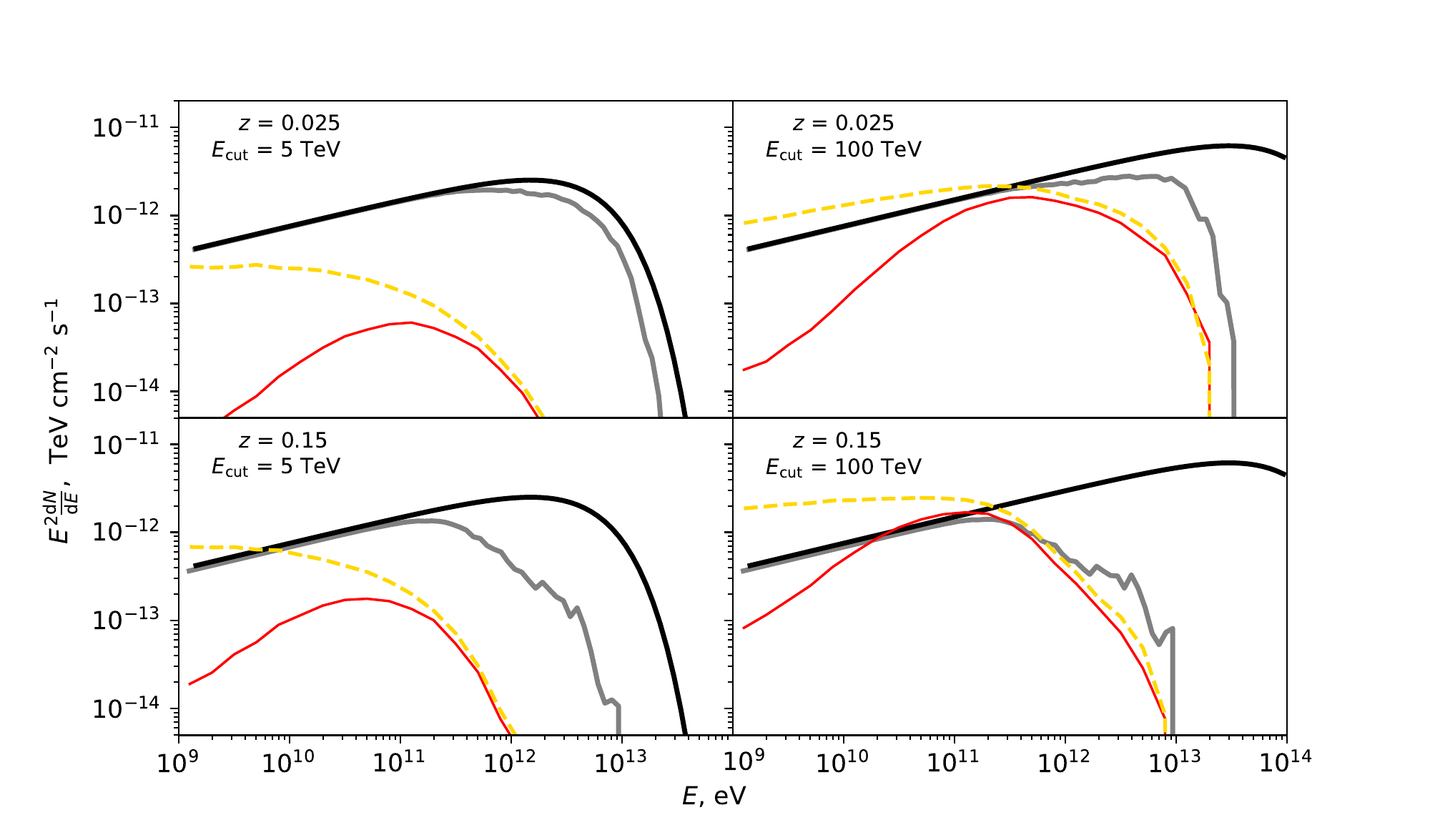}
        \caption{ Primary and secondary \gr\ spectra for the ``only voids'' magnetic field profile (only regions with $B \leq 10^{-12}$~G are taken into account).  Different subplots represent different parameters of the source. Solid black and gray lines are the intrinsic and observed primary source spectra. Dashed yellow lines show the secondary \gr\ flux for a zero line of sight magnetic field. Solid red lines show the secondary \gr\ flux with a nonzero magnetic field in the voids. \label{fig:spec_only_voids}}
\end{figure*}

In order to separately study the effect of bubbles and voids on the cascade component, we constructed three alternative magnetic field profiles for each LOS. One profile contained only voids: in all voxels where $B \geq 10^{-12}$~G, we set $B=0$~G. The second profile contained only bubbles: the magnetic field in all voids with $B<10^{-12}$~G and in the first bubble around the source AGN was set to $B=0$~G. The last profile contained only the AGN source bubble: the magnetic fields everywhere outside the first bubble were set to $B=0$~G. To avoid the effect of random fluctuations of the magnetic field and stabilize the procedure of determining the size of the first bubble better, we defined the boundary of the first bubble as the point from which the magnetic field strength is below $10^{-12}$~G for at least $L_\mathrm{b}=300$ kpc. We verified that if we replaced $L_\mathrm{b}$ with 100, 500, or 1000 kpc, the size of most of the first bubbles did not change, and if it did, the changes were insignificant.

We propagated \gr s from the source to the observer for each magnetic field profile and calculated the secondary \gr\ signal along the LOS. For LOSs containing only the first bubble, potentially more significant fluctuations of the secondary flux may arise because all primary photons are emitted in the same direction. In reality, the primary \gr s are emitted in a cone with an opening angle $\alpha_\mathrm{jet}$ so that they sense different directions in the bubble. The bubble properties in various directions are different (size and magnetic field profile). In our analysis, we neglected this difference, and the bubble radius changes probably only insignificantly on the scale of the jet opening angle.

We explored several possible spectral models of the AGN source. The intrinsic \gr\ spectrum of the source was always a cutoff power law $\mathrm{d}N/\mathrm{d}E\propto E^{-\Gamma}\exp(-E/E_\mathrm{cut})$. The slope of the power law is $\Gamma=1.7$. In contrast, $E_\mathrm{cut}$ was a free parameter. Another free parameter was the distance to the observer $D_\mathrm{s}$ which sets the redshift of the source $z$. We scanned over parameter space choosing $E_\mathrm{cut}$ in the 1-100~TeV range and $0.025<z<0.15$. Cosmological parameters were set to $H_0 =$ 70 km/(s Mpc), $\Omega_\mathrm{m}=0.3,$ and $\Omega_\mathrm{\Lambda}=0.7,$ and the EBL model was fixed to that of \citet{Gilmore:2011ks}. 

We propagated \gr s until they reached the sphere with radius $r=D_\mathrm{s}$. Although all primary particles are emitted in the same direction, we were able to model primary photon emission into a jet with an opening angle $\alpha_\mathrm{jet}$  by recording the \gr s that hit the part of the sphere lying inside the cone with the opening angle $\alpha_\mathrm{jet}=5^\circ$, whose axis coincides with the direction from the source to the observer. For the spectral plots presented in the next section, we retained only the \gr s that arrived at the position of the telescope at an incidence angle within the telescope PSF, which we assumed to be  $\theta_\mathrm{psf}=0.3^\circ$.

\section{Results}

\begin{figure*}
    \includegraphics[width=\textwidth]{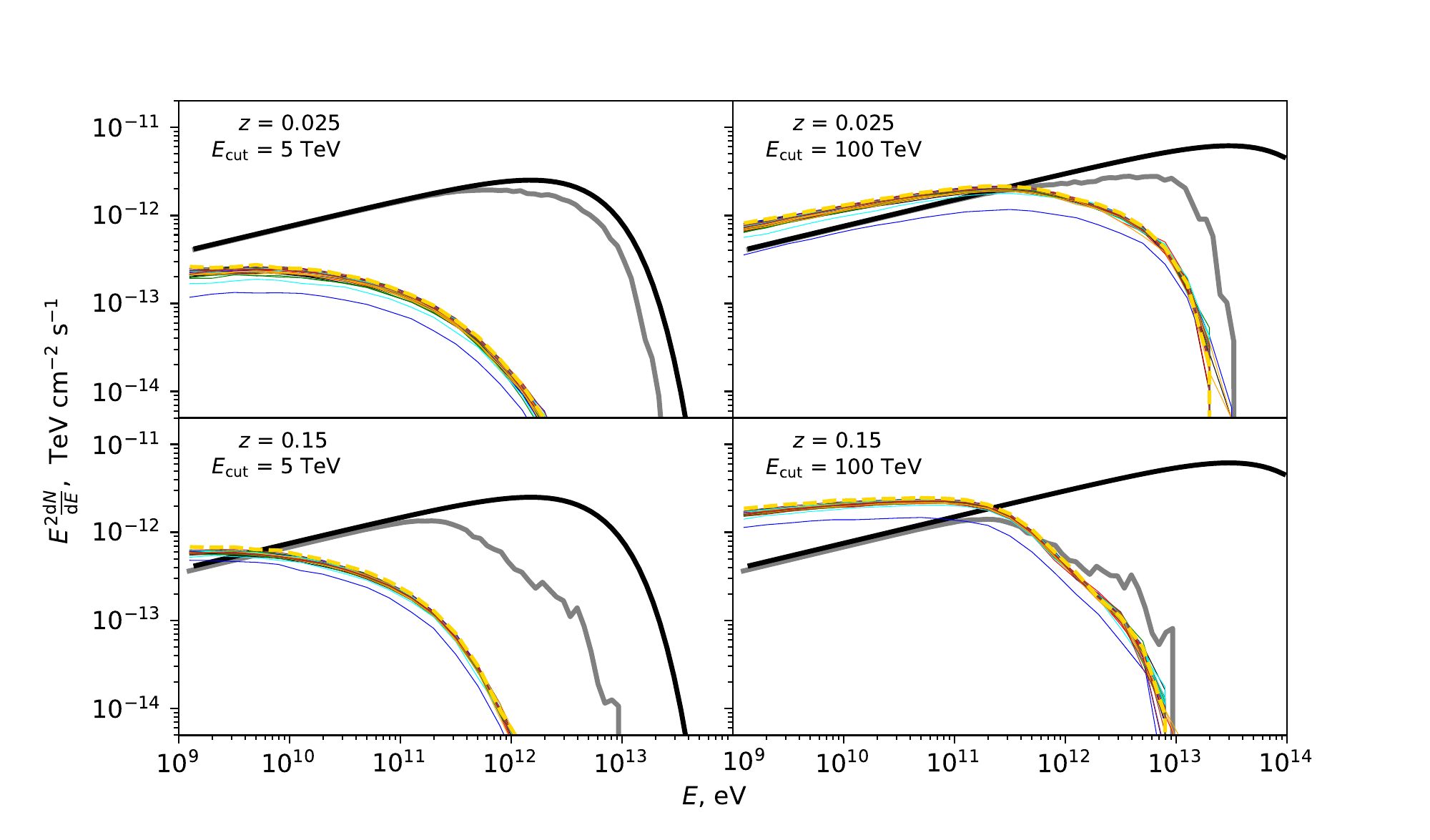}
        \caption{ Same as Fig.~\ref{fig:spec_only_voids}, but for the ``only bubbles'' magnetic field profiles (only regions with $B<10^{-12}$~G are taken into account). Colored solid lines indicate different LOSs.} \label{fig:spec_all_other_bubbles}
\end{figure*}

\begin{figure*}
    \includegraphics[width=\textwidth]{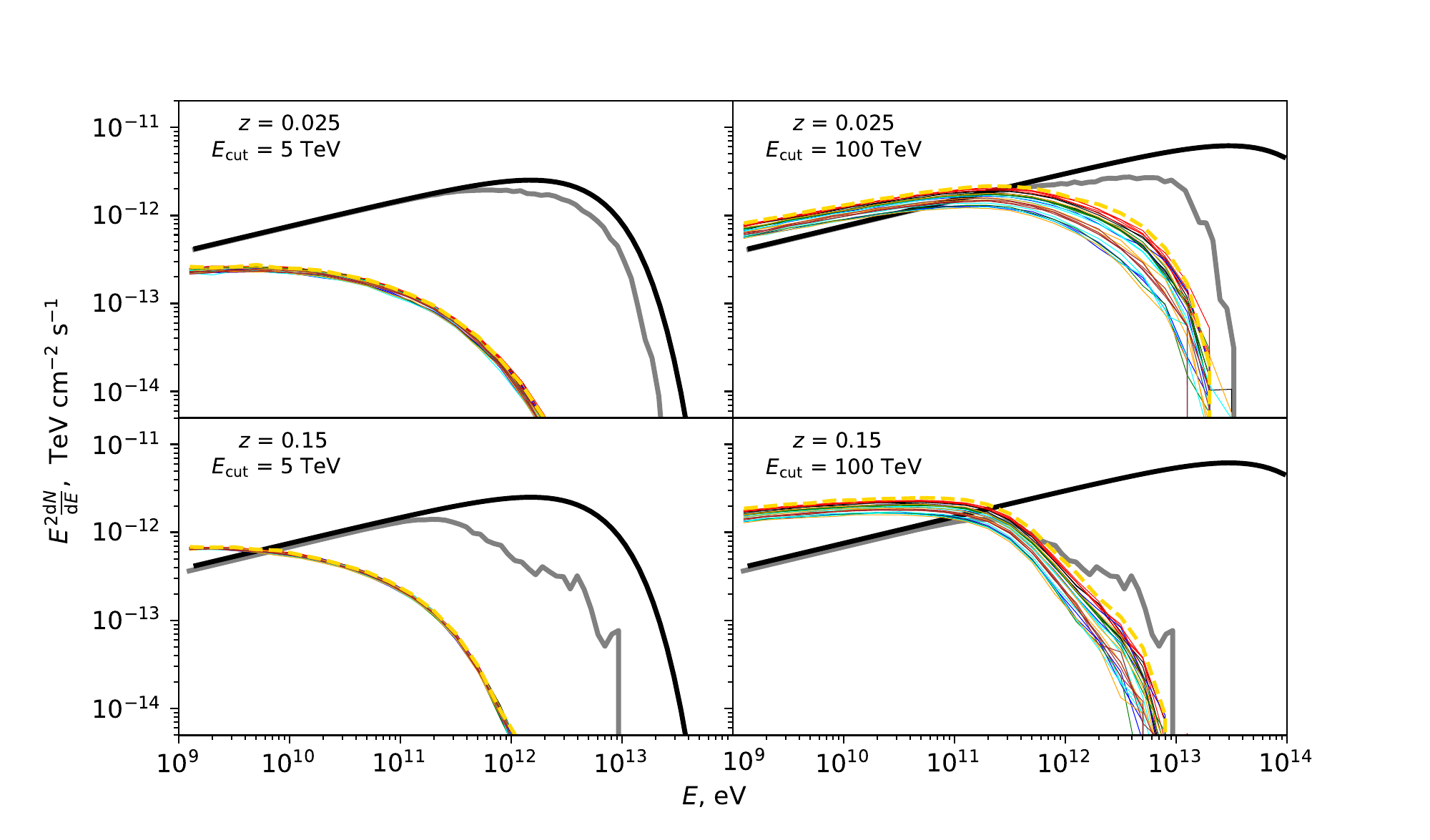}
        \caption{ Same as Figs. \ref{fig:spec_only_voids} and \ref{fig:spec_all_other_bubbles}, but for the ``only host bubble'' magnetic field profile (only the region with $B>10^{-12}$~G around the source galaxy is taken into account).  \label{fig:spec_only_host_bubble} }
\end{figure*}

The simulated spectra of the primary AGN source and secondary \gr\ emission are shown in Figures \ref{fig:spec_only_voids}, \ref{fig:spec_all_other_bubbles}, and \ref{fig:spec_only_host_bubble}. They represent three types of magnetic field profiles and four different choices of parameters of the AGN source: a nearby source with low energy cutoff, a nearby source with high energy cutoff, a distant source with low energy cutoff, and a distant source with high energy cutoff. 

The three sets of figures show that the effect of the void IGMF on the secondary \gr\ signal (Fig.~\ref{fig:spec_only_voids}) is clearly different from the effect of the bubble magnetic fields (Figs. \ref{fig:spec_all_other_bubbles} and \ref{fig:spec_only_host_bubble}). The volume-filling IGMF outside bubbles produces complete suppression of the secondary \gr\ emission at $E<100$ GeV. As the initial conditions of TNG100 contain a constant magnetic field with a strength of $B=10^{-14}$~G, this part of the IGMF, unaffected by galactic dynamos, has a long correlation length and $B>10^{-15}$~G almost everywhere (see Fig.~2 in~\citet{Garcia:2020kxm}). Therefore, such an effect on the secondary \gr\ emission is not unexpected, based on the analytical estimates of \citet{Neronov:2007zz,Neronov:2009gh}.

The magnetized bubbles produce a qualitatively different effect on the secondary \gr\ flux. Strong ($B>10^{-12}$~G) magnetic fields in the bubbles sufficiently deflect the charged particles that are created in the bubbles and therefore completely suppress their secondary \gr\ flux at the detector throughout its energy range. This changes the overall normalization of the secondary emission, but affects its shape only very little. 

This qualitative difference between the effects of void IGMF and bubbles holds for the whole range of properties of the primary \gr\ sources. Only the strength of the overall flux suppression by the bubbles depends on the source parameters. 

\begin{figure*}
    \includegraphics[width=\textwidth]{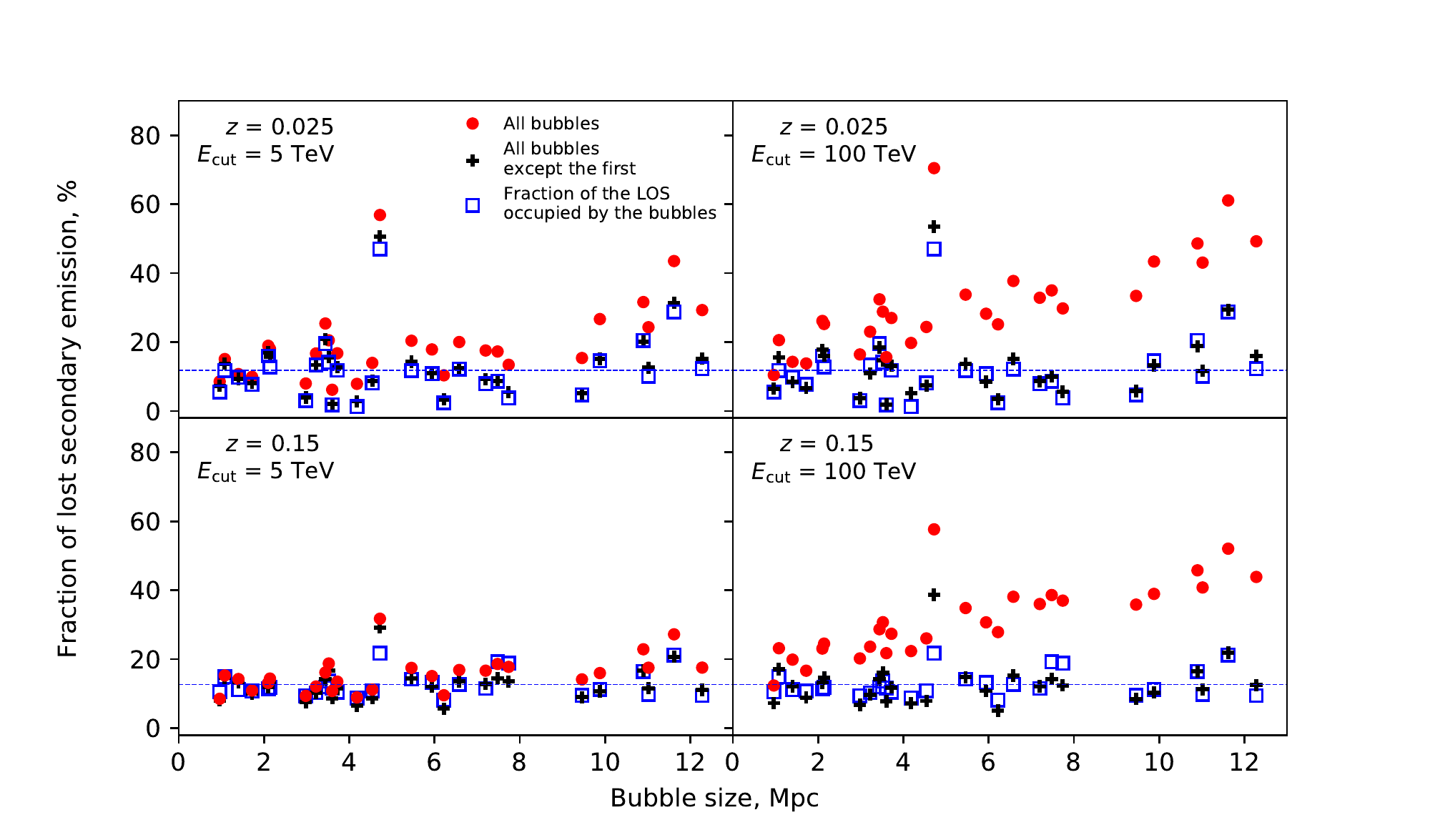}
        \caption{ Fraction of the secondary \gr\ flux in the energy range 1-10 GeV that is removed by the bubbles as a function of the size of the host AGN bubble. Red circles show the flux fractions that are removed by all bubbles. Black crosses and blue squares show the flux fractions that are removed by all bubbles, except for the host AGN bubble and a linear fraction of the LOS that is occupied by these bubbles. The dashed blue horizontal line marks the average linear fraction of the LOS that is occupied by all bubbles, except for the host AGN bubble.} \label{fig:lost_fraction_bubble_size}
\end{figure*}

The level of secondary flux suppression clearly depends on the length of the bubbles along the line of sight from the sources toward the observer. Fig.~\ref{fig:lost_fraction_bubble_size} shows the lost secondary flux in percentage points for all the simulated LOSs. Within the TNG100 feedback model, the magnetized bubbles typically remove some 10\% of the secondary \gr\ flux if the source intrinsic \gr\ spectrum has a high energy cutoff at relatively low energy $E_\mathrm{cut}\sim 5$~TeV. However, for a fraction of the LOSs, the pass of the photons through the initial bubble happens to be large (above 10 Mpc), and loss of the secondary emission for these cases is two to three times larger. 

The average flux suppression increases by up to 20\%-50\% for hypothetical sources with hard intrinsic spectra extending up to $E_\mathrm{cut}=100$~TeV. Even though there is currently no observational evidence for the existence of such extreme AGNs, we show the result for this case to highlight the dependence of the bubble effect on the intrinsic properties of the primary \gr\ source.  

The secondary flux suppression by the bubbles is accumulated in the AGN host bubble and all other bubbles along the line of sight. The relative importance of the ``host'' and ``all other'' bubbles also depends on the properties of the primary \gr\ source. This is illustrated in Fig.~\ref{fig:lost_fraction_bubble_size}, which also shows the percentage of the lost secondary flux as a function of the bubble type. 

The suppression by ``all other'' bubbles except for the AGN host bubble is at the level 5-20\%, and it does not depend on the distance to the source or on the cutoff energy. In this case, the suppression factor is well described by the fraction of the LOS length occupied by the magnetized bubbles. 

One singular LOS in Fig.~\ref{fig:lost_fraction_bubble_size} was affected more strongly by the bubbles (50-60\%), despite a modest (5 Mpc) pass in the initial bubble. The reason for this phenomenon is clear from  Fig.~\ref{fig:LOSexamples}: in this case, the LOS passes not only through the initial bubbles around the source, but also through a large filamentary system of magnetized bubbles that occupy more than half of the first 100 Mpc of this LOS. Although this seems to be a rare combination of the location of the source and the orientation of the LOS with respect to the filamentary system of bubbles, this possibility can not be dismissed for any particular source (at least according to the IllustrisTNG model).

The AGN host bubble has practically no effect on the secondary source flux if the cutoff energy $E_\mathrm{cut}$ is low. However, it has a strong impact on the signal from sources with high cutoff energy. In this case, the mean free path $D_\mathrm{mfp}(E_\mathrm{cut})$ of the gamma-rays with energy $E_\mathrm{cut}$ becomes smaller than the size of the first bubble, so that a significant fraction of primary \gr\ flux is absorbed inside this AGN host bubble. This significantly increases the percentage of lost secondary emission, as shown in Fig.~\ref{fig:lost_fraction_bubble_size}. The effect is more substantial for higher cutoff energies, and it is slightly weakened for further away sources through the additional contribution to the secondary flux from longer propagation to the observer. For nearby sources with high cutoff energy, additional losses in the first bubble can exceed 30\%. 

\begin{figure*}
    \includegraphics[width=\textwidth]{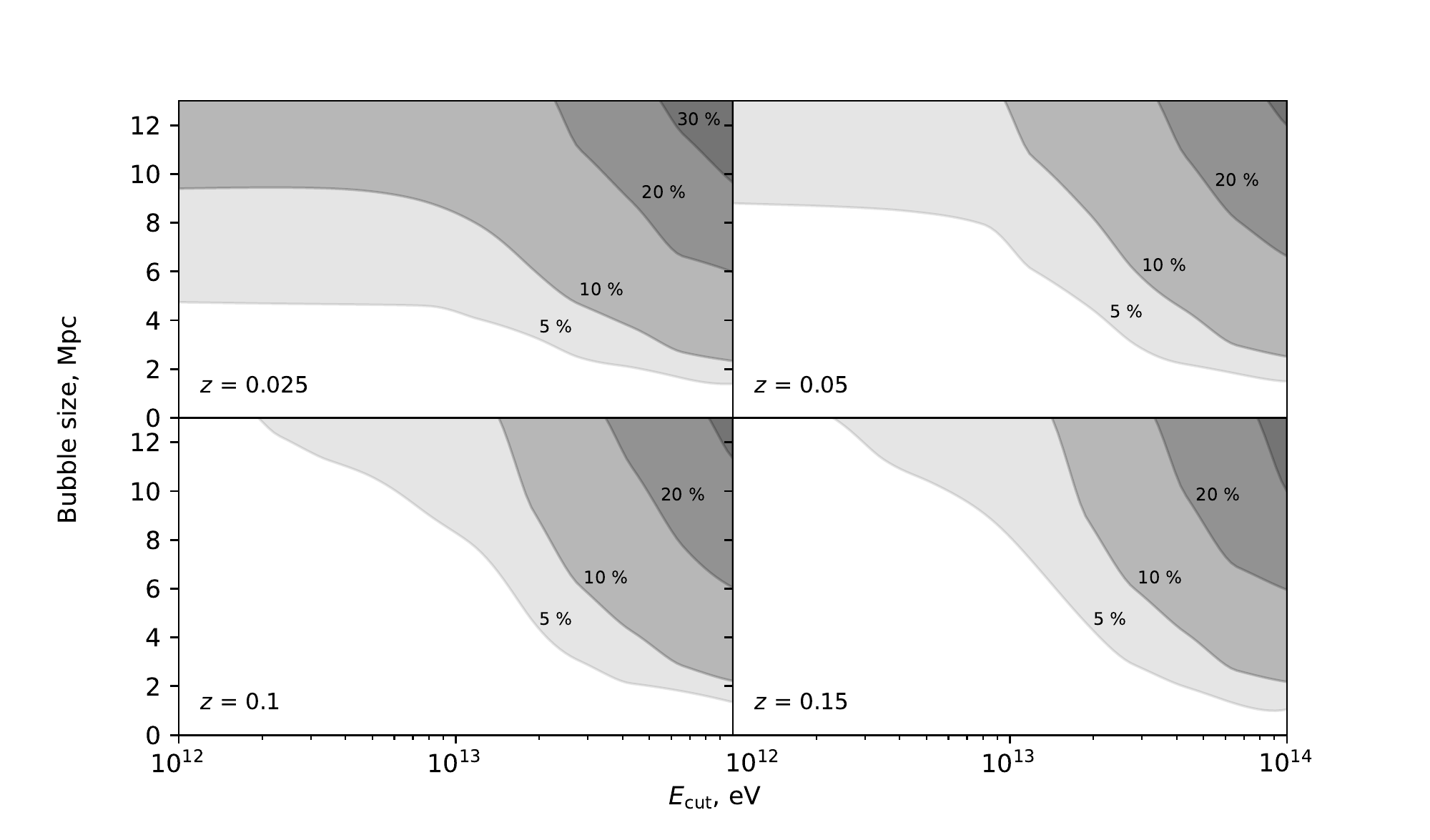}
        \caption{ Fraction of the secondary flux in the energy range 1-10 GeV that is removed by the host AGN bubble as a function of cutoff energy and size of the bubble. \label{fig:lost_fraction_cutoff}}
\end{figure*}

The range of the primary source parameters for which the losses in the first bubble are perceptible is shown in Fig.~\ref{fig:lost_fraction_cutoff}. Two source classes are clearly visible. Nearby sources with low cutoff energies are practically unaffected by the host AGN bubbles. This is true as long as the mean free path of gamma-rays $D_\mathrm{mfp}(E_\mathrm{cut})$ exceeds the distance to the observer $D_\mathrm{s}$ , so that the suppression factor is equal to the fraction of the LOS occupied by the bubble. This explains the ``horizontal branch'' of the contours in  Fig.~\ref{fig:lost_fraction_cutoff}.  As soon as the mean free path of the primary \gr s becomes shorter than the distance to the source, the fraction of the flux that is removed by the AGN host bubble is given by the ratio of the bubble size to the primary \gr\  mean free path $D_\mathrm{mfp}(E_\mathrm{cut})$. This effect is visible toward the higher energy ``inclined'' branch of the contours in  Fig.~\ref{fig:lost_fraction_cutoff}. This is an important conclusion that can be drawn from Fig.~\ref{fig:lost_fraction_cutoff}. The presence of magnetized bubbles in the LSS produces only a mild effect on the secondary \gr\ flux at the 10\% level for all reasonable AGN source parameters. In this respect, the bubbles have to be taken into account as an additional source of systematic uncertainties in the analysis of the IGMF in voids. 

\section{Discussion}

The results presented above show that an account of the magnetic fields produced by the baryonic feedback on the LSS modifies model calculations of the secondary \gr\ flux from the electromagnetic cascade that develops along the line of sight. The strength of the effect depends on the parameters of the primary \gr\ source.

In the specific feedback model of the IllustrisTNG simulation, the effect is always at the level of $\sim 10\%$ for a wide range of AGN primary source parameter choices: cutoff energies below $\sim 30$~TeV, and redshifts within $z\sim 0.15$. In this case, the possible presence of magnetized bubbles along the line of sight has to be considered as an additional source of systematic uncertainty of the overall level of the secondary flux. This systematic uncertainty does not affect the spectral shape of the secondary \gr\ flux. In general, it does not affect the timing and extended emission profiles of the signal either. This implies that previously developed analysis methods leading to constraints on the strength and correlation length of the void IGMF remain valid even in the presence of the magnetized bubbles. 

It is interesting to note that the qualitative difference of the effect of the bubble and void IGMF on the \gr\ signal can be used to separate the two effects in the observational data. The bubbles completely randomize the directions of electrons and positrons in regions of high magnetic fields. In this respect, they produce the same effect, independently of the energy of electrons and positrons. In contrast, the void IGMF randomizes the directions of only low-energy electrons and positrons. If the primary source flux is well constrained (this suggests that a high-statistics measurement of the attenuated point-source flux at the highest energy is available), the total secondary flux level can be reliably estimated. An overall energy-independent suppression of the secondary flux can therefore be measured. This can provide an estimate of the fraction of the line of sight occupied by the magnetized bubbles. The void IGMF is measurable through a different effect of the existence of extended or delayed emission below a characteristic energy at which the IGMF is able to sufficiently deflect electrons and positrons away from the line of sight. 

The separation between the bubble and IGMF effects on the secondary \gr\ flux becomes less clear when the time delay of the cascade signal is very long or when sources have a very high cutoff energy and hard intrinsic spectrum. In the first case, the intrinsic luminosity of the source cannot be constrained. In the second case, the mean free path of the primary \gr s can be comparable to the size of the bubble around the AGN host galaxy. Suppression of the secondary flux can be strong in this case (see Fig.~\ref{fig:lost_fraction_bubble_size}). This introduces a huge systematic uncertainty in the model of the secondary \gr\ flux. This uncertainty can preclude an estimate of the IGMF, especially if the statistics of the \gr\ data on the extended or delayed signal is poor. 

\section{Conclusions}

We have explored the effect of the baryonic feedback on the LSS in the form of outflows from galaxies that pollutes the intergalactic medium with magnetic fields. This process can spread the magnetic field in the voids of the LSS and thus preclude measuring the primordial magnetic fields that possibly reside in the voids. 

Our analysis of the specific baryonic feedback model implemented in IllustrisTNG cosmological simulations shows that the feedback generally does not strongly affect the void magnetic fields. The galactic outflows from magnetized bubbles collectively have a small volume filling factor. 

We have explored the effect of the magnetized bubbles on the \gr\ signal from distant AGN. This \gr\ signal is used for the measurements of magnetic fields in the voids. We characterized the effect of the magnetized bubbles on the signal and found that they typically produce an overall suppression of the secondary flux. Within the IllustrisTNG baryonic feedback model, the effect is generally small, at the level of $\sim 10\%$ for representative parameters of the primary AGN \gr\ sources. However, we found one LOS in our sample for which effect is about $60\%$ because the first 100~Mpc are massively polluted by magnetic bubbles. When a single source is studied, this possibility can therefore not be excluded. Robust measurements have to rely on a large sample of sources. In this case, the effect of magnetized bubbles on the \gr\ signal (energy-independent suppression of the secondary flux) is readily distinguishable from the effect of the void magnetic fields (which completely suppresses the secondary \gr\ flux below some characteristic energy).

The effect of outflow-driven magnetized bubbles can be taken into account in the analysis as a systematic uncertainty of the secondary \gr\ flux at the level of 10\% on average (and up to $60\%$ for a single source). This error has to be taken into account in the error budget of the analysis. This systematic uncertainty depends on the high energy cutoff of the primary source spectrum. If the cutoff energy is in the $E_\mathrm{cut}=100$ TeV range, a dedicated analysis of the possible properties of the AGN host bubble has to be performed as part of the IGMF-related data analysis.

\begin{acknowledgements}
Work of D.S. and A.N. has been supported in part by the French National Research Agency (ANR) grant ANR-19-CE31-0020, work of A.K. was supported in part by Russian Science Foundation grant 20-42-09010. A.K.'s stay in the APC laboratory was provided by the  ``Vernadsky'' scholarship of the French embassy in Russia. K.B. and A.B. are supported by the European Research Council (ERC) Advanced Grant ``NuBSM'' (694896). A.S. is supported by the FWF Research Group grant FG1.
\end{acknowledgements}

\bibliographystyle{aa}
\bibliography{refs.bib}

\end{document}